RESEARCH　　　　　　　　　　　　　　　　　　　　　　　　　　　　　　　　　　　　　Open Access

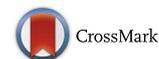

# The role of pedagogical tools in active learning: a case for sense-making

Milo Koretsky[1*] , Jessie Keeler[1], John Ivanovitch[2^] and Ying Cao[1]

## Abstract

**Background:** Evidence from the research literature indicates that both audience response systems (ARS) and guided inquiry worksheets (GIW) can lead to greater student engagement, learning, and equity in the STEM classroom. We compare the use of these two tools in large enrollment STEM courses delivered in different contexts, one in biology and one in engineering. Typically, the research literature contains studies that compare student performance for a group where the given active learning tool is used to a control group where it is not used. While such studies are valuable, they do not necessarily provide thick descriptions that allow instructors to understand how to effectively use the tool in their instructional practice. Investigations on the intended student thinking processes using these tools are largely missing. In the present article, we fill this gap by foregrounding the intended student thinking and sense-making processes of such active learning tools by comparing their enactment in two large-enrollment courses in different contexts.

**Results:** The instructors studied utilized each of the active learning tools differently. In the biology course, ARS questions were used mainly to "check in" with students and assess if they were correctly interpreting and understanding worksheet questions. The engineering course presented ARS questions that afforded students the opportunity to apply learned concepts to new scenarios towards improving students' conceptual understanding. In the biology course, the GIWs were primarily used in stand-alone activities, and most of the information necessary for students to answer the questions was contained within the worksheet in a context that aligned with a disciplinary model. In the engineering course, the instructor intended for students to reference their lecture notes and rely on their conceptual knowledge of fundamental principles from the previous ARS class session in order to successfully answer the GIW questions. However, while their specific implementation structures and practices differed, both instructors used these tools to build towards the same basic disciplinary thinking and sense-making processes of conceptual reasoning, quantitative reasoning, and metacognitive thinking.

**Conclusions:** This study led to four specific recommendations for post-secondary instructors seeking to integrate active learning tools into STEM courses.

## Introduction

Our program recently interviewed faculty candidates for an open position. During the interview process, each candidate was asked to conduct a 20-min teaching demonstration. One candidate, a tenured associate professor from a large, public research university, had regularly taught core courses. He enthusiastically stated that he had incorporated active learning into his courses and asked to use clickers as part of the demonstration. In the first 15 min, the candidate delivered a transmission-oriented PowerPoint presentation on heat transfer. This lecture portion was followed with a multiple-choice clicker question. In the question, the instructor provided a short word problem related to the material and an equation that he had just presented. He asked the audience to select which variable in the equation was the unknown among a list of variables that appeared in the

* Correspondence: milo.koretsky@oregonstate.edu
^Deceased
[1]School of Chemical, Biological, and Environmental Engineering, Oregon State University, Corvallis, OR 97331, USA
Full list of author information is available at the end of the article equation. All the information needed to answer the question was provided in the question stem, and it could clearly be answered simply by variable identification, independently of understanding the material that had been presented earlier. More insidiously, this question reinforced an undesirable schooling practice of many students—searching a source to find an appropriate equation and variable matching. When asked his objective for incorporating clickers into his course, the candidate stated, "I just want to make sure my students are awake."

Motivated by the overwhelming evidence that demonstrates the effectiveness of active learning over traditional lecture in science, technology, engineering, and mathematics (STEM) courses (e.g., Freeman et al. 2014; Hake 1998; Prince 2004), many instructors are seeking to transform their classroom practice to incorporate active learning (Borrego et al. 2010; Felder and Brent 2010). However, as illustrated by the vignette above, these instructional practices can be taken up in a range of ways, and the instructor's conception of learning is critical. We believe that the faculty member above chose to employ clicker technology in a way that made sense to him and that less productive enactments of active learning can be logical interpretations of research studies that predominantly focus on the effectiveness of a practice relative to its absence. In this qualitative, comparative case study, we investigate the ways experienced instructors choreograph such activity in their courses to produce learning and thereby seek to provide a complementary lens for instructors to productively implement active learning in their course.

We call the clicker applied in the vignette above an active learning *tool*. Tools are used in instruction to place students in an environment where they interact in intentional ways with the content and with other students and the instructor to promote learning. Tools can be technology-oriented like the clicker technology above or pedagogically oriented like the guided inquiry worksheets we describe below and often combine aspects of both orientations. Researchers who study the efficacy of these tools typically compare student performance for a group where the given tool is used to a control group where it is not used. Such research focuses on the tool's effect (what learning gains does it produce?) and how to use the tool (what do instructors need to learn about to use it?). In many cases, incorporation of tools provides evidence of increased learning outcomes. However, this avenue of research implicitly can lead to undesired translations to practice based solely on considerations of procedure about how to use the tool, as illustrated in the vignette above. Investigations on the intended student thinking processes (not performance gains) using these tools are largely missing. In the present article, we fill this gap by foregrounding the intended thinking and sense-making processes of such technological and pedagogical tools.

We compare the use of active learning tools in two STEM courses delivered in different contexts (one in biology and the other in engineering). Both courses use the same two tools: audience response systems (ARS) and guided inquiry worksheets (GIW). They are both taught by instructors experienced with active learning pedagogies and recognized as high-quality and innovative instructors by their peers and students. We are interested in how implementation of these tools varied between courses and in identifying threads common to both. We focus on the intended student sense-making and thinking processes as the instructors integrate the tools into their courses. By sense-making, we follow Campbell et al. (2016) to mean that learners are "working on and with ideas—both students' ideas (including experiences, language, and ways of knowing) and authoritative ideas in texts and other materials—in ways that help generate meaningful connections" (p. 19). Our goal is not to compare learning gains in these two courses in order to claim one instructor's implementation strategy works better than the other. Rather through analysis of the similarities and differences in the course design and practices, we seek to provide a lens into how active learning comes to "life," and to provide instructors productive ways to think about how they can best integrate active learning into their classroom learning environment.

We ask the following research questions. In two large-enrollment undergraduate STEM courses in different disciplinary contexts:

1. What types of thinking and sense-making processes do instructors intend to elicit from students during their use of ARS questions? During their use of GIW questions? What are the similarities and differences between the intended uses of these tools in the two courses studied?
2. In what ways do the intended sense-making processes that are elicited through the use of the ARS and GIW tools align with the instructors' broader perspectives and beliefs about the instructional system for their courses?

## Background

To situate this study, we first provide an overview of the research on ARS and GIW tools. We then describe the thinking and sense-making processes on which we will focus to understand the ways that the instructors in this study use the tools in concert and how they integrate them to achieve outcomes of instruction.

### Audience response systems as tools

ARS, such as clickers, have been used increasingly in post-secondary STEM classroom to allow instructors to shift large classes from a transmission-centered lecture mode into active learning environments (Freeman et al. 2014; Hake 1998; Prince 2004). Typically, the instructor provides the class a multiple-choice conceptual question, and each student in the class responds by selecting an answer on a device. In some cases, students are also asked to provide written explanations justifying their answer selection (Koretsky et al. 2016). Aggregate responses are available for the instructor to display to the class in real time. Often, students are asked to discuss answers in small groups, in a whole class discussion, or both (Nicol and Boyle 2003).

ARS tools elicit live, anonymous responses from each individual student allowing students in the class to answer new questions in a safe manner free from judgment of peers and the instructor (Lantz 2010). In addition, real-time availability of the answer distribution can provide immediate and frequent feedback and allows for adaptable instruction. Based on student responses, instructors can modify class discussion and activity to meet learning needs that are more representative of the entire class rather than just a few vocal students. However, instructors also have concerns about incorporating ARS in classes, including fear about covering less content, less control in the student-centered classroom, and the time and effort needed to learn the technology and develop good questions (Caldwell 2007; Duncan 2005; MacArthur and Jones 2008).

The research literature on ARS use has focused broadly on both student engagement and student learning. Synthesis of individual research studies has shifted from more descriptive review papers (Caldwell 2007; Duncan 2005; Fies and Marshall 2006; Kay and LeSage 2009; MacArthur and Jones 2008) to more systematic meta-analyses (Castillo-Manzano et al. 2016; Chien et al. 2016; Hunsu et al. 2016; Nelson et al. 2012) that use common metrics and statistical methods to relate the characteristics and findings of a set of studies that are selected from explicit criteria (Glass et al. 1981). In general, researchers report ARS tools promote student engagement by improved attendance, higher engagement in class, and greater interest and self-efficacy (Caldwell 2007; Kay and LeSage 2009; Hunsu et al. 2016) and also suggest that anonymity increases engagement (Boscardin and Penuel 2012; Lantz 2010).

Research on student learning with ARS tools often takes an interventionist approach, comparing classes or sections where instructors use the ARS to those that only lecture (Chien et al. 2016; Castillo-Manzano et al. 2016) or, occasionally, contrasting ARS technology with the same in-class questions delivered without using ARS technology, such as by raising hands, response cards, or paddles (Chien et al. 2016; Elicker and McConnell 2011; Mayer et al. 2009). Learning gains are often measured from instructor-developed assessments, such as in-class exams (Caldwell 2007), but more robust psychometric instruments such as concept inventories have also been used (Hake 1998). Results generally, but not always, show improved outcomes (Hunsu et al. 2016; Boscardin and Penuel 2012). These reports also acknowledge the relationship between ARS use, and learning is complex (Castillo-Manzano et al. 2016). Many factors have been suggested to influence it, such as the depth of the instructor's learning objectives (Hunsu et al. 2016), testing effects (Chien et al. 2016; Lantz 2010; Mayer et al. 2009), and the extent of cognitive processing (Beatty et al. 2006; Blasco-Arcas et al. 2013; Mayer et al. 2009; Lantz 2010) and social interactions (Blasco-Arcas et al. 2013; Chien et al. 2016; Penuel et al. 2006).

In summary, there is large and growing body of literature that has examined the use of ARS tools in STEM courses. These studies suggest that they are effective in eliciting student engagement and learning, especially in large classes.

### Guided inquiry worksheets as tools

GIW are material tools that guide students through inquiry learning during class. In general, inquiry learning seeks to go beyond content coverage and engage students in the practices of doing science or engineering, e.g., investigating a situation, constructing and revising a model, iteratively solving a problem, or evaluating a solution (National Research Council 1996; de Jong and Van Joolingen 1998). However, inquiry can be challenging for students since it requires a set of science process skills (e.g., posing questions, planning investigations, analyzing and interpreting data, providing explanations, and making predictions) in addition to content knowledge (National Research Council 2011; Zacharia et al. 2015). In guided inquiry, instructional scaffolds provide support to help students effectively engage in scientific practices around inquiry (Keselman 2003; de Jong 2006). There have been several pedagogies that embody inquiry learning which range from less guided approaches like problem-based learning (PBL) to more guided approaches like process-oriented guided inquiry learning (POGIL, Eberlein et al. 2008).

Guided inquiry learning activities are pedagogically grounded and guide students through specific preconceived phases of inquiry (Pedaste et al. 2015). For example, both POGIL (Bailey et al. 2012) and peer-led team learning (PLTL, Lewis and Lewis 2005, 2008; Lewis 2011) are designed to guide students through a three-phase learning cycle (Abraham and Renner 1986): (i) the *exploration phase* where students search for patterns and meaning in data/models; (ii) the *invention phase* to align thinking around an integrating concept; and (iii) the *application phase* to extend the concept to new situations. Similarly, pedagogically grounded inquiry-based learning activities (IBLAs, Laws et al. 1999; Prince et al. 2016) contain three phases intended to produce a cognitive conflict that elicits students to confront core conceptual ideas: (i) the *prediction phase* where students make predictions about counter-intuitive situations; (ii) the *observation phase* where they observe an experiment or conduct a simulation; and (iii) the *reflection phase* which consists of writing about the differences and connecting to theory.

GIW are commonly used as tools to provide carefully crafted *key* questions that guide students through the conceived phases of inquiry during class (Douglas and Chiu 2009; Eberlein et al. 2008; Farrell et al. 1999; Lewis and Lewis 2008). Lewis (2011) describes GIW as typically containing from six to 12 questions that vary between a conceptual and procedural nature. Questions often progress in complexity (Bailey et al. 2012; Hanson and Wolfskill 2000). First, they might ask students to explore a concept, thereby activating their prior knowledge. Then, they ask students to interact with models and develop relationships and finally elicit students to apply the learned concepts to new situations, thereby generalizing their knowledge and understanding. When inquiry is centered on observations of a phenomenon, GIW provide a tool for students to write down both their initial predictions and their observations, thereby producing a written record that they must reconcile (Prince et al. 2016).

Similar to findings on ARS tools, the research literature indicates that guided inquiry pedagogies promote engagement (Abraham and Renner 1986; Bailey et al. 2012), learning (Abraham and Renner 1986; Lewis 2011; Prince et al. 2016; Wilson et al. 2010), and equity (Lewis and Lewis 2008; Lewis 2011; Wilson et al. 2010) in the STEM classroom.

### Thinking and sense-making processes

Our study situates the intersection of pedagogical strategies and content delivery in the intended thinking and sense-making processes of students as they engage in active learning tasks. We take a constructivist perspective of learning (National Research Council 2000; Wheatley 1991) that views new knowledge results from students' restructuring of existing knowledge in response to new experiences and active sense-making. This restructuring process is effectively carried out through interactions with other students in groups (Chi and Wylie 2014; Cobb 1994). From this perspective, a key aspect of instruction then becomes to create and orchestrate these experiences.

Educators can design and implement learning activities in ways to cultivate productive thinking and sense-making processes while delivering course content. As emphasized in STEM 2026, a vision for innovation in STEM education, "[a]lthough the correct or well-reasoned answer or solution remains important, STEM 2026 envisions focus on the process of getting to the answer, as this is critical for developing and measuring student understanding." (Tanenbaum 2016, p. 33).

### Conceptual reasoning

In this study, conceptual reasoning refers to the reasoning processes where individuals and groups draw on foundational disciplinary concepts and apply them in new situations (National Research Council 2000, 2013). Elements of conceptual reasoning include (but are not limited to) identifying appropriate concepts when analyzing a new problem or situation, understanding those concepts and their relationship to the context, and applying the concepts to solve problems or explain phenomena. (Russ and Odden 2017; Zimmerman 2000). Facility with concepts and principles has been identified as a feature of thinking that distinguishes disciplinary experts from novices (National Research Council 2000).

Researchers have suggested several changes from traditional instructional design that better align with developing students' conceptual reasoning (e.g., Chari et al. 2017; National Research Council 2000, 2013). First, instruction should shift to more in-depth analysis of fewer topics that allows focus and articulation of key, crosscutting concepts (National Research Council 2013). In doing so, the curriculum must provide sufficient number of cases to allow students to work with the key concepts in several varied contexts within a discipline (National Research Council 2000). Second, classroom activities should provide students with opportunities to practice conceptual reasoning on a regular basis. Instructors can prompt this practice by asking students questions which require conceptual reasoning. They should also hold students accountable for such reasoning by participating in discussion, modeling thinking, and steering students away from rote procedural operations towards conceptual reasoning (Chari et al. 2017).

### Quantitative reasoning

Quantitative reasoning addresses analysis and interpretation of numerical data and application of quantitative tools to solve problems (Grawe 2016), as well as mathematical sense-making—the process of seeking coherence between the structure of the mathematical formalism and the relations in the real world (Dreyfus et al. 2017; Kuo et al. 2013). Quantitative reasoning has been recognized as a key learning outcome for twenty-first century college graduates (Association of American

Colleges and Universities 2005). Quantitative reasoning at the college level includes processes such as translating between verbal, graphical, numeric, and symbolic representations; interpreting measured data or mathematical models; and using mathematical methods to numerically solve problems (Engelbrecht et al. 2012; Mathematical Association of America 1994; Zimmerman 2000). The word "reasoning" suggests the synthesis of quantitative concepts into the greater whole (Mathematical Association of America 1994) and emphasizes the *process* of performing the analysis instead of merely the *product* that results from it. In the context of upper division college courses, quantitative reasoning tends to be even more sophisticated where "the connections between formalism, intuitive conceptual schema, and the physical world become more compound (nested) and indirect" (Dreyfus et al. 2017, p. 020141-1).

Quantitative reasoning reflects the incorporation of mathematical knowledge and skills into disciplinary contexts (Mathematical Association of America 1994). In science and engineering, quantitative reasoning can include making sense of measured data and connecting it to physical phenomena (Bogen and Woodward 1988) or developing mathematical models that predict and generalize (Lehrer 2009; Lesh and Doerr 2003). Thus, the use of mathematics extends beyond the procedures and algorithms that students sometimes take as synonymous with the field. Researchers claim that mathematical sensemaking is possible and productive for learning and problem solving in university science and engineering courses (e.g., Dreyfus et al. 2017; Engelbrecht et al. 2012).

In addition, conceptual reasoning and quantitative reasoning are intertwined in disciplinary practice and should be cultivated in tandem. Researchers have identified several ways that conceptual reasoning aids in science and engineering problem solving, including conceptualization for finding the equations to mathematically solve the problem, checking and interpreting the result after the equation is solved, and the processes working through to the solution (Kuo et al. 2013). Zimmerman (2000) points out that domain-specific concepts, i.e., "thinking within the discipline," and domain-general quantification skills (e.g., evaluating experimental evidence) "bootstrap" each other and, when conducted together, lead to deeper understanding and richer disciplinary knowledge and skills.

In a culture that often focuses on and rewards procedural proficiency, it can be challenging to engage students in quantitative reasoning (Engelbrecht et al. 2012). Active learning strategies can help (Grawe 2016). Strategies include emphasizing accuracy relative to precision, asking students to create visual representations of data or translate between representations, asking students to communicate about their quantitative work, and setting assignments in an explicit, real-world context (Bean 2016; Grawe 2016; MacKay 2016).

### Metacognitive thinking

Metacognition often refers to "thinking about thinking" or "second-order-thinking," the action and ability to reflect on one's thinking (Schoenfeld 1987). Research evidence suggests that metacognition develops gradually and is as dependent on knowledge as experience (National Research Council 2000). Ford and Yore (2012) argued that critical thinking, metacognition, and reflection converge into metacognitive thinking and can improve the overall level of one's thinking.

Vos and De Graaff (2004) claimed that active learning tasks, such as working on projects in engineering courses, do not just require metacognitive knowledge and skills but also encourage the development of the learners' metacognitive thinking. Based on several decades of research literature, Lin (2001) concluded that there are two basic approaches to developing students' metacognitive skills: training in strategy and designing supportive learning environments.

Veenman (2012) pointed out three principles for the successful instruction of metacognitive thinking: (1) Metacognitive instruction should be embedded in the context of the task; (2) learners should be informed about the benefit of applying metacognitive skills; and (3) instruction and training should be repeated over time rather than being a one-time directive. When designing STEM curriculum in an integrated way, an issue that becomes central is determining the aspects of metacognition and the context in which the aspects should be taught (Dori et al. 2017).

## Methods

To answer our research questions, we use a comparative case study of two STEM courses implementing both an ARS and GIW. Data for this study were collected within a larger institutional change initiative whose goal is to improve instruction of large-enrollment STEM courses across disciplinary departments through implementation of evidence-based instructional practices (Koretsky et al. 2015). Through multiple data sources, we seek to provide a thick description (Geertz 1994) of how and why instructors use these active learning tools in large classes.

### Case selection

We selected courses based on the regular use of ARS and GIW tools as part of classroom instruction. In addition, we sought courses in different disciplinary contexts and department cultures since the instructors would more likely show variation in tool use. Based on these criteria, we identified courses in biology, in engineering, and in a third STEM discipline. Based on the instructor's willingness to cooperate, we ended up investigating the biology and the engineering course in this study. Both are taught in the same public university, have large student enrollments, and are required courses for students majoring in the respective disciplines. Both instructors had experience using these tools for several terms prior to our investigation and were identified by peers and students as excellent educators.

The biology course, *Advanced Anatomy and Physiology*, is the third course in an upper division sequence series required for biology majors. Prerequisites for the course included the completion of the preceding courses in the year-long sequence and concurrent or previous enrollment in the affiliated lab section. The enrollment was 162 students in the term studied. The engineering course, *Material Balances*, is the first course of a required three-course sequence for sophomores majoring in chemical, biological, and environmental engineering. The enrollment was 307 students in the term studied.

### Data collection and analysis
Data sources include a series of interviews with each instructor, classroom observations and instructional artifacts, and student response records to ARS questions.

#### Instructor interview protocol
We conducted four semi-structured interviews with each instructor over the span of three academic years. The first (year 1) and fourth (year 3) interviews were reflective interviews that probed the instructors' general teaching practices and instructional beliefs. They included questions about department and college duties, interactions with other faculty regarding teaching and learning, perceptions about successful students, and responses to the larger change initiative. The second and third interviews (year 2) focused specifically on the ARS and GIW questions, respectively, applied to a specific delivery of the instructor's course and are described in more detail below. All interviews were audio-recorded.

The interviews on the ARS and GIW questions sought to elicit the instructor's understandings of the questions they assigned and their rationale for assigning them, the reasons and purposes of why they used the tool (ARS or GIW), and how they used these tools in the greater context of their courses. To investigate the intended types of student thinking processes for each active learning tool, we asked the instructors to write out their solutions to the questions following a think-aloud protocol (Ericsson 2006). For these interviews, each instructor was interviewed by a researcher who had deep domain knowledge in the course content under study. The researcher provided the instructors with hard copies of selected ARS questions (Interview 2) and GIW questions (Interview 3) and asked them to write their responses on them, which were then collected for analysis. To gain insight into the instructor's perception of the questions when he or she positioned him- or herself as a learner, we began each interview with the following prompt: "I want you to imagine you're looking at these questions from a student's perspective, and I want you to talk through how you would answer each one." The think-aloud portion was followed by a reflective portion. After the instructors worked through all the questions, they were directed through the question set a second time with the prompt: "What was your rationale when assigning this question?"

#### Selection of ARS questions for think-aloud interview
For the interview with the biology instructor, we decided it was feasible, given the brevity of the biology ARS questions, to have the instructor work through all 31 ARS questions delivered in the Friday POGIL sessions in an hour-long interview. The engineering ARS questions required more time than the biology questions, and we decided it was not feasible to expect the instructor to work through 31 ARS questions in an hour-long interview. We chose a subset of diverse questions that were representative of the whole set. Criteria for choosing questions included the difficulty of the question (determined by the percent correct scores from the students), the topic of each question (based on the topics outlined in the course syllabus), question clarity, and how the percent correct changed if peer instruction was involved. Following these criteria, we selected 15 of the 31 questions to present to the engineering instructor.

#### Selection of GIW questions for think-aloud interview
During interviews, we asked the instructors to engage with GIW questions from selected weekly worksheets. We selected a single guided inquiry worksheet for the think-aloud interview with the biology instructor. The worksheet focused on parameters of vascular function. We chose this worksheet based on the instructor's input that it was representative of the type of worksheets students would encounter during a guided inquiry session for her course. For the engineering course, the instructor expressed two approaches to worksheet development. One was more conceptually based and one was more computational. We chose two worksheets, one for each approach. The first worksheet was selected because it was the first worksheet that applied the concept of a material balance (conservation of mass); this concept was then incorporated into almost all of the subsequent guided inquiry worksheets in the class. The second worksheet was a later-term worksheet that asked students to use Excel to perform calculations and then answer questions involving quantitative and qualitative answers. We first asked the instructors to answer the GIW questions as if they were students encountering the worksheet for the first time and subsequently asked them to explain their rationale for placing each question on the worksheet.

### Interview analyses

We transcribed all the interviews verbatim and analyzed interviewee responses using emergent coding. For the think-aloud interviews, this process was used to identify the general intended thinking processes that occurred as instructors worked and talked through ARS and GIW questions from the perspective of the students in their course. We examined each ARS or GIW question response the instructors gave and identified individual steps. We assigned code to each step describing what the instructor was doing at that step. Then, we recognized sets of individual steps from different questions belong to more general categories that represented the broader type of thinking and sense-making processes described in our theoretical framework (i.e., conceptual reasoning, quantitative reasoning, and metacognitive thinking). In such cases, we grouped them into a more general code category. For example, codes such as "use graphical information to qualitatively explain situation," "relate information to physical representation," and "identify relationships between variables" were grouped in the more general code category "conceptual reasoning." Similarly, codes such as "develop equations to describe phenomena," "rearrange equation to identify relationships between variables," and "perform calculations" were grouped together in the more general code category "quantitative reasoning." The approach of identifying categories from specific thinking processes led to a reliable coding process. By grouping, we were able to develop a general set of codes that connect the data with our theoretical framework. We could then compare thinking processes (i) between the courses representing different disciplines and (ii) between different pedagogical tools within each course. Table 1 provides the final set of code categories for the intended thinking processes during the think-aloud interviews including a description of the code and a sample interview quote for each code.

For the reflective interviews, we sought to relate the ways the instructors used the ARS and GIW tools to their priorities about what students would learn and their conceptions about how students learn and the role of teaching in learning (Thompson 1992). Code categories were determined as follows. One researcher coded the interview transcript initially and developed a set of emergent categories based on elements of the instructional system that the instructors mentioned. The

**Table 1** Code definitions and example responses from for ARS questions

| Code | Description | Example of thinking process |
| --- | --- | --- |
| Immediate recall | Answering the question from work completed during the immediate class session | "So to answer this, I think I would look to my worksheet, so whatever that model is. So that model with the, oh it's graphs as I recall, so model with graphs and then I would reference my conversation that I had." |
| Recognize concept(s) | Explicitly identifying the main concept(s) involved in a question | "Okay, I'm thinking, again, degrees of freedom, but we've got multiple phases here and so we need to use Gibbs phase rule…" |
| Compare available answers for best choice | Reviewing available multiple choice answers to decide which answer made the most sense | "I think that I would have to say okay, you know, we just went through these four answers from this worksheet, and, you know, we eliminated some of these options basically as being the wrong answer." |
| Select information from question | Referring to the question to obtain needed information to work towards the answer | "I'm gonna go back to my problem and look at umm you know, what are the things I'm given umm is there something that makes sense in terms to base this calculation on." |
| Conceptual reasoning | After identifying a concept, using fundamentals to reason through to an answer (e.g., using a graphical representation or an equation to think through how the variables related to one another) | "…think about the mass flow rate and use that to say that the mass in and out is gonna be the same. So I'm going to, yeah, focus on that concept of mass conservation here." |
| Quantitative reasoning | Developing equations to describe what was happening in the question and also possibly using a numerical calculation. | "And in this case I want to do a material balance on C… I'm just gonna write it out, in minus out plus generation minus consumption equals zero…" |
| Metacognitive thinking | Reflecting on the context that the question is asked or the meaning or reasonableness of an answer | "Ultimate aim of the process is to produce dry crystalline, sodium bicarbonate so what did I think this process was for, looks like we're actually trying to make the solid phase as opposed to reducing the concentration in the liquid phase." |
| Recall lecture information/prior knowledge | Using information presented in lecture or other course resources to make progress | "Alright so it's uhh it's telling me that umm it's going to be example problem similar to one that we worked in lecture." |

Koretsky et al. International Journal of STEM Education (2018) 5:18 Page 8 of 20research team then met and reconciled the categories. This process resulted in the following subset of categories that were relevant to this study: instructional scaffolding, constructivism, social interactions, formative assessment, summative assessment, and sense-making.

For both think-aloud and reflective interviews, a single researcher with appropriate expertise performed the coding. A subset of the interview transcripts (20–25%) was then coded by a second researcher to ensure reliability (Cohen's kappa = 0.80 [think-aloud], 0.84 [reflective]).

### Other measures

We used several other data sources to triangulate our interview data analysis and interpretation. Both classes were observed several times using the Teaching Dimensions Observation Protocol (TDOP, Hora et al. 2013). While the report of TDOP codes for each course is beyond the scope of this article, the observation process allowed researchers to get familiar with the course structure and the context in which the ARS and GIW active learning tools were used. The observations were supported with instructional artifacts including the course syllabus, all the ARS questions and GIW activities, and the week's lecture notes for the analyzed GIW activities. Student responses for ARS questions were collected through the web-based platform that each instructor used to assign questions and receive student responses. The response data were used to verify our interpretation of the different intents of the instructors in their use of ARS questions.

## Context of tool use

In this section, we describe how each instructor situates the use of ARS and GIW tools within the activity structure of their courses. This description is based on analysis of the course observations and course artifacts and triangulated with the reflective interviews.

Table 2 provides a summary of the differences in the context and use of the active teaching tools between the biology and engineering courses. We unpack these enactments when we present the results.

### Biology course

The biology course met three times per week with more traditional lecture periods on Monday and Wednesday and an active learning period of POGIL guided inquiry sessions on Friday. ARS questions were used in most class periods, whereas the GIW tool was used only in the Friday POGIL class periods. Over the 10-week term, students answered a total of 98 ARS questions, 31 of them during the POGIL GIW sessions. They completed nine GIW in total. An average of around 110 out of 162 students attended GIW sessions.

Figure 1 shows an example of a biology ARS question after it was delivered in class. ARS questions were presented to students as PowerPoint slides, and the students answered using clickers. The instructor typically displayed the distribution of students' choices on the screen shortly thereafter and briefly commented on the correct answer. Students answered between two to five questions per class period. Our think-aloud interview with the biology instructor focused on the ARS questions delivered on the Friday POGIL sessions.

The instructor used a GIW tool to facilitate student activity during the guided inquiry sessions on Fridays. A typical Friday session (50 min) consisted of an introduction (2 min), student group work using the GIW and facilitated by undergraduate learning assistants (15–25 min), ARS questions and group discussion (5–10 min), continued student group work (10–15 min), and wrap up (2–5 min). During these sessions, the classroom was divided into seven territories with a learning assistant assigned to each territory. The instructor assigned students to groups of three which were maintained for the duration of the term. Students collaboratively worked on a GIW answering an average of 19 worksheet questions during each session. The GIWs engaged students in interpreting models of an underlying concept as well as providing data that students used to answer questions. Some worksheets also contained "extension questions" that students engaged in outside of class or in class if they finished the worksheets ahead of other groups. Extension questions typically consisted of two types, classified by the instructor as (i) *big picture questions*, consisting of open-ended questions to stimulate discussion and look at larger concepts, and (ii) *clinical correlation questions*, situating concepts in clinical applications.

**Table 2** Alignment across different elements of the two courses

| | Biology | | | Engineering | | |
|---|---|---|---|---|---|---|
| | Frequency | Size | Purpose | Frequency | Size | Purpose |
| Lecture | Twice a week, 50 min | 165 students | Content is introduced, and students engage in ARS questions and discussions. | Twice a week, 50 min | 300 students | Content is introduced, and example problems are solved. |
| ARS questions | Three times a week | 165 students | "Check in" with students to see if they are correctly interpreting content or ask students to report their understanding of worksheet questions just completed. | Once a week, 50 min | Approx. 150 students/ section | Strengthen conceptual understanding by building on topics from lecture through ARS questions about new situations. |
| GIW questions | Once a week, 50 min | 165 students | Guide students through biological models, enabling them to engage with the content using disciplinary thinking. | Once a week, 50 min | Approx. 30 students/ section | Scaffolded GIW questions reinforce problem solving skills/processes instituted in lecture. |

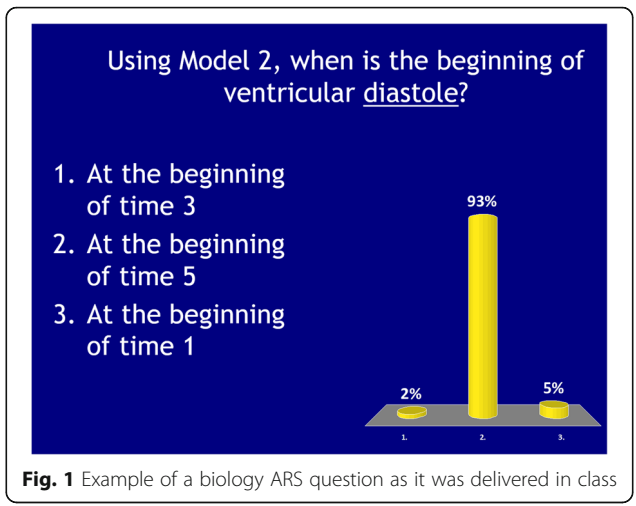

Fig. 1 Example of a biology ARS question as it was delivered in class

Figure 2 shows the first part of the GIW tool that we used for the think-aloud interview. The worksheet contained three different models; each model was followed by 2 to 9 questions students answered based on information from the models. The worksheet ended with six "extension questions" that the students answered outside of class time. Although these extension questions required more critical thinking than the previous questions, they still typically referenced the models contained in the worksheet.

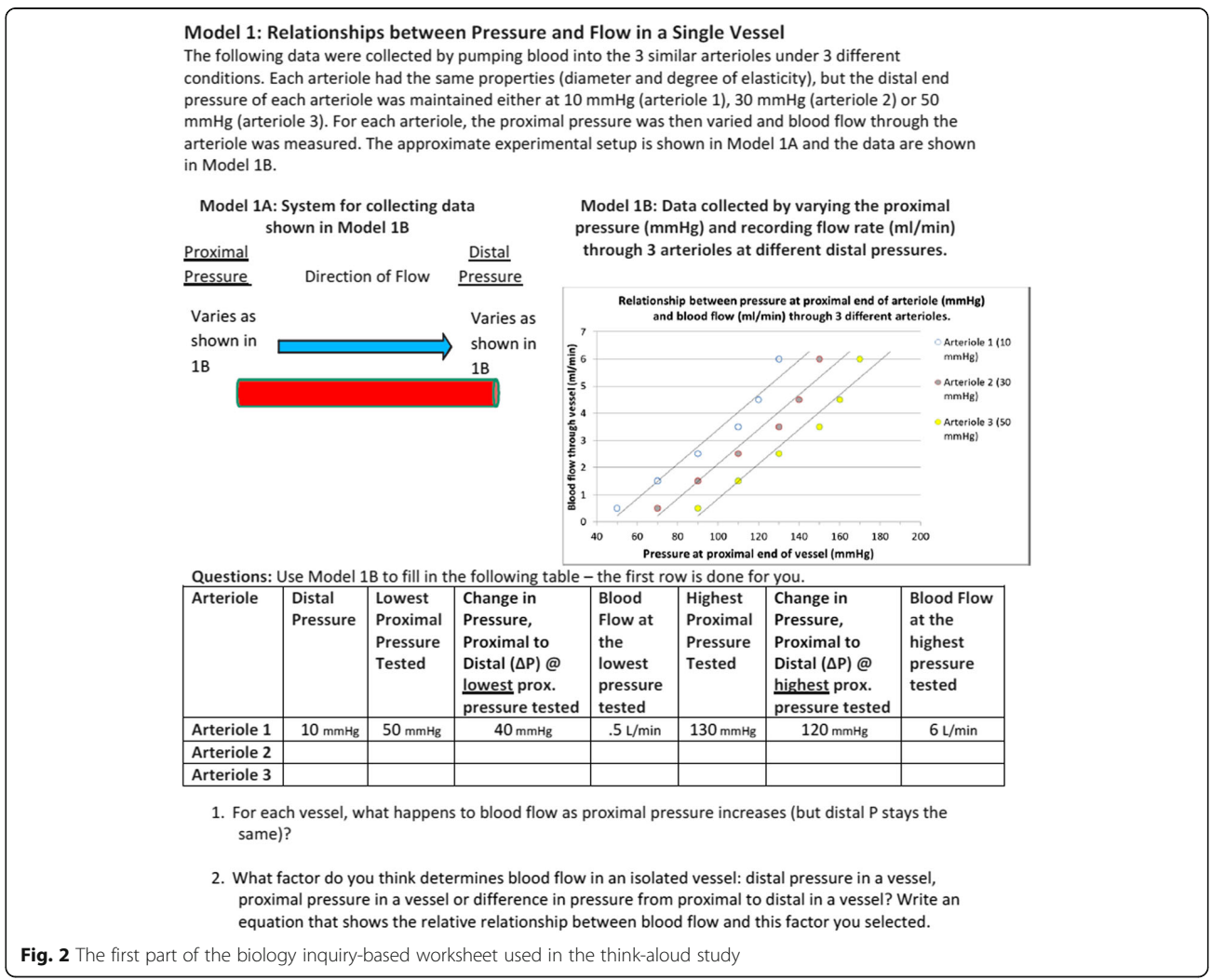

Fig. 2 The first part of the biology inquiry-based worksheet used in the think-aloud study

### Engineering course

The engineering course has two 50-min lectures on Monday and Friday attended by all students. ARS questions were delivered during the 50 min Wednesday sessions. GIWs were used during the 50-min Thursday sessions. Over the course of the 10-week term, the students answered a total of 31 ARS questions and completed nine GIW in total. An average of 285 out of 307 students responded to questions during the ARS sessions, and almost all students attended the GIW sessions.

In the Wednesday sessions, students were divided into two sections of approximately 150 students and ARS questions were delivered via the *AIChE Concept Warehouse* (Koretsky et al. 2014) by the primary instructor to each section separately. Figure 3 shows an example of an engineering ARS question. Students responded to the questions using their laptops, smartphones, or tablets, and the answers were stored in a database. Along with their multiple-choice answer, students were asked to provide a written justification of their answer as well as a confidence score between one and five to indicate how sure they were of their answer. For about half of the ARS questions (15 questions), the engineering instructor used an adaption of the peer instruction (PI) pedagogy (Mazur 1997) where students first answered a posted question individually, then discussed their answers with their neighbors, and then re-answered the same question again (including written justification and confidence). For the 16 non-PI questions, the instructor displayed the responses and discussed the question with the whole class after the students answered individually.

On Thursdays, students attended smaller "studio" sessions of approximately 30–36 students where they completed GIW activities. Most studios were facilitated by a graduate teaching assistant (GTA) who would sometimes briefly (1–2 min) introduce the topic. For most of the studio period, students spent their time actively working. Some worksheets involved an introduction section that was to be completed individually, while others solely contained group work. Studio groups contained three or four students. Group members were kept the same for 5 weeks, after which students were assigned to new groups by the GTA for the remainder of the 10-week term. GTAs were coached to respond to student questions with subsequent

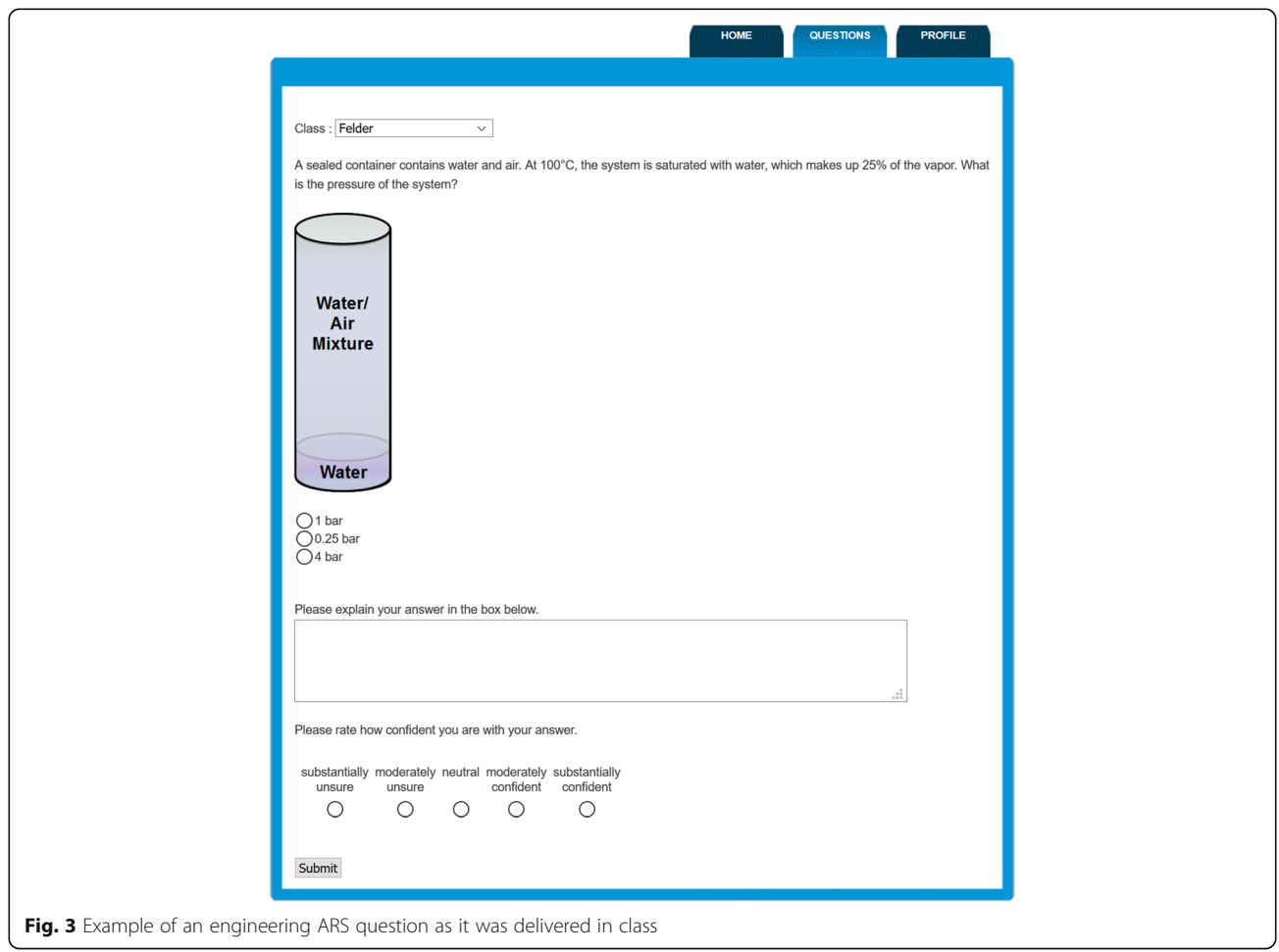
**Fig. 3** Example of an engineering ARS question as it was delivered in class

guiding questions, as opposed to providing students with direct answers to the worksheet questions. Students were given 50 min to work on the worksheets which were collected at the end of each studio session.

An example of part of a GIW used in the engineering think-aloud interview is shown in Fig. 4. Most worksheets involve a main problem that includes a problem statement with relevant information students need to use to solve the worksheet. The problem is then broken down into different steps that students complete to help them reach an answer. These steps are modeled after problem solving processes introduced in lecture and may include questions which require students to read and interpret the problem statement, to draw and label a diagram that represents what is happening in the problem, to formulate and solve appropriate equations, and to relate the problem statement to real-world engineering scenarios.

## Results

In the following sections, we answer our two research questions with evidence from our data analysis.

### Answer to RQ1: What are the instructors' intended thinking processes during use of ARS and GIW tools? What are the similarities and differences between instructors?

Our answer to Research Question 1 is based on analysis of think-aloud interviews for the ARS questions and the GIW questions, triangulated by student responses of ARS questions, researcher in-class observations, and the instructors' reflective interviews.

### Thinking processes elicited from the ARS and GIW questions

Table 3 shows the percentage of questions identified for each of the categories of thinking process during the think-aloud interviews (see Table 1 for category definitions). Results from the ARS questions are shown in the left column for each course, and results from the GIW questions are next to them to the right. The majority of ARS questions in the biology course focused on immediate recall (75%). While the biology instructor expected students to be able to recognize a concept in 25% of the questions, there was no evidence that the ARS questions were intended to evoke further conceptual reasoning. In contrast, the engineering instructor rarely sought to elicit immediate recall (7%), but rather to provide students experiences where they needed to select information from questions (47%) and recognize concepts (73%) to prompt conceptual reasoning (73%). In addition, the majority of the questions also included elements of quantitative reasoning (60%).

During her think-aloud interview, the biology instructor showed a wide range of intended scientific thinking processes in responding to the GIW questions, including conceptual reasoning (42%), quantitative reasoning (42%), and metacognitive thinking (32%). These

---

A container holds 10.0 kg of a saturated solution of $NaHCO_3$ at $60°C$. The solubility of $NaHCO_3$ in water at $60°C$ is 16.4 g/100.0 g $H_2O$. The temperature of the solution is reduced to $27°C$ to crystallize 500 g of $NaHCO_3$. Find the final amount of solution in kg and the final concentration of $NaHCO_3$ in solution in g $NaHCO_3$/g solution.

Steps:

1. <u>Read and understand the problem</u>:

    a. Is the process continuous or batch?

    b. What is to be accomplished by this process?

    c. What does the term saturated mean in this problem?

2. <u>Draw a diagram and label it appropriately</u>. Be sure to include numerical values and units for known quantities and symbols for unknown quantities. Write down the units for the unknown quantities.

Fig. 4 The first part of one of the engineering inquiry-based worksheets that was used in the think-aloud study

Table 3 Percent of questions identified for coded thinking processes for ARS questions and GIW questions in biology (BIO) and engineering (ENGR) courses. Definitions of codes are presented in Table 1

| Thinking process | BIO | | ENGR | |
| --- | --- | --- | --- | --- |
| | ARS questions (31 questions) | GIW questions (19 questions) | ARS questions (15 questions) | GIW questions (20 questions) |
| Immediate recall | 75% | 47% | 7% | 45% |
| Recognize concept(s) | 25% | | 73% | |
| Compare available answers for best choice | 22% | | 20% | |
| Select information from question | | 95% | 47% | 65% |
| Conceptual reasoning | | 42% | 73% | 5% |
| Quantitative reasoning | | 42% | 60% | 60% |
| Metacognitive thinking | | 32% | | 10% |
| Recall lecture information/prior knowledge | | 21% | | 85% |

processes were not sequestered but rather the instructor integrated each one around thinking about models of the fluid dynamics of vascular function. The worksheet tended to be "stand alone" with information usually found in the question (95%) rather than intending students to recall information from lecture (21%). The biology worksheet is self-contained in the sense that most of the information students need to complete the worksheet is provided via the models. In contrast, during the think-aloud interview, the engineering instructor intended students to spend the majority of time engaged in quantitative reasoning (60%) with only a small amount of time in conceptual reasoning (5%). There is also less metacognitive thinking in these activities (10%) than that in the biology course. In addition, the engineering instructor intended students to reference previous knowledge and information presented in lecture (85%) to a much larger degree than the biology instructor.

While the uses of ARS and GIW tools in each course is distinct and different, inspection of Table 3 shows that in either course by the time students completed that week's active learning activity, they were intended to significantly engage around a key disciplinary topic in two of the aspects of thinking and sense-making: conceptual reasoning and quantitative reasoning. Thus, the "coverage" of the topic extends beyond declarative content and patterns of problem solving. Rather, it emphasizes productive ways to think and reason in the discipline. The biology instructor had more explicit intended metacognitive thinking than the engineering instructor (32 vs. 10%). However, for all 31 ARS questions, the engineering instructor had students rate their confidence (see Fig. 3), so while he did not allude to metacognitive thinking as much during the think-aloud interviews, there was some of this type of thinking built into the technology tool.

### Student ARS performance

We next present student performance data from the ARS questions for each course. These data show differences in types of questions asked and implementation and reflect differences in intended thinking processes discussed previously. Figure 5 shows the percentage of students who answered correctly for each question when the ARS questions were delivered in class. Results from the 31 ARS questions delivered during Friday POGIL sessions in the biology course are shown chronologically with red diamonds (labeled BIO). Students performed well in general averaging 89.3% correct (solid red line) with a standard deviation of 10.5. In the engineering course, students' initial responses are shown with solid dark blue circles (ENGR pre-PI). They averaged 58.5% correct (solid dark blue line) on these questions with a standard deviation of 20.2. For the 15 questions where the engineering instructor used the PI pedagogy, the post-PI question results are shown by powder blue circles. Their average correct was 80.0% (powder blue line) with a standard deviation of 14.8%. Of the questions where PI was used, scores increased by an average of 18.0%, showing benefit of peer discussion, although there were two questions where scores significantly decreased (Q12 and Q21).

The nature of the ARS questions is clearly different in these two courses: the engineering questions were more difficult and took more class time, and when peer instruction pedagogy was used, they were asked twice. These differences both reflect the context where a weekly class period was dedicated to the ARS questions in engineering and the intended thinking processes during the think-aloud interviews with the instructors shown in Table 3. We next explore the reflective interviews with the instructors to see how these uses align with their conceptions of how these tools fit into their courses to produce learning.

### Instructor perceptions of the ARS and GIW questions

In this section, we present excerpts and analysis of the four reflective interviews with each instructor: the year 1

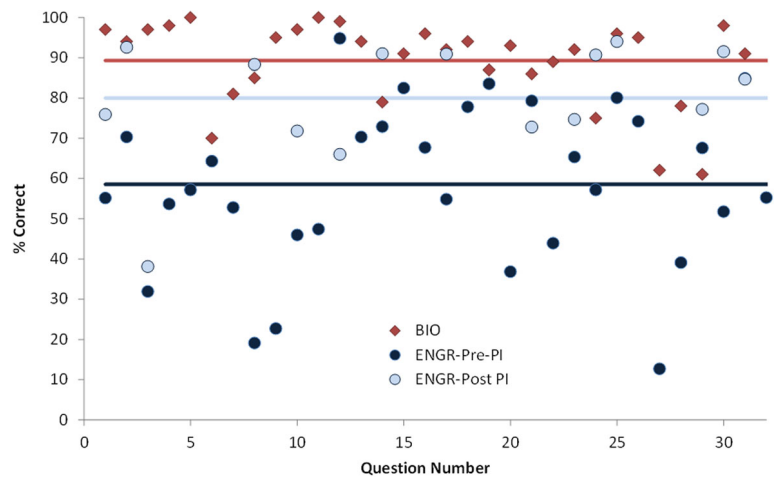

Fig. 5 ARS question student performance data

general interview (labeled g-pre) and the year 3 general interview (g-post) focused on more general questions about their instructional practices and beliefs while the year 2 post think-aloud interviews (post think-aloud ARS or GIW) specifically addressed the instructors' intent in using these tools.

**ARS questions** The reflective interviews corroborate the identified differences between the courses in the ways the ARS questions engaged students. In the biology course, ARS questions were used mainly to assess if students were correctly interpreting and understanding worksheet questions or to ask students to recall the material recently introduced in the worksheet. Each of the biology ARS questions was applied once. When questioned about her rationale behind designing ARS questions, the biology instructor acknowledged that they can be a helpful tool to use in large classes.

> Biology Instructor (post think-aloud ARS): ...and so you know the value of clicker questions in rooms of greater than 100 people is... in a really big class, I can't see their sheets, and so I don't know what they're thinking. And *it's really useful to check in with them in that way*. [italic added for emphasis]

She alluded to the role of ARS questions as "concept checking," and pointed out that she regularly uses them to "check in" with students to ensure they are engaged and following along in class.

The engineering course presented ARS questions that afforded students the opportunity to apply learned concepts to new scenarios towards improving students' conceptual understanding. When the engineering instructor was asked about how he wants students to be engaged while solving ARS questions, he explained that he wanted to push students beyond procedural problem solving:

> Engineering Instructor (post think-aloud ARS): I guess trying to get them to start to create that knowledge structure in their head. That there are certain conventions and there are certain cues that are gonna help them bend those problems into, you know, help them find a solution...*Trying to provide cues that are similar to things they're gonna see in other parts of the class, homework and exams and so forth, to get them to hone in on those specific concepts* and then in some cases manipulate or examine at a level that they're not gonna get just by plugging and chugging into those equations. [italic added for emphasis]

This line of thinking tied into comments in the year 1 general reflective interview where the engineering instructor referred several times to the ability to conceptually reason by identifying a concept and applying it to a new situation, such as in the following excerpt:

> Engineering instructor (g-pre): If you can understand the fundamentals, the fundamental concepts that are governing a process then, you know, if you start to change all these other things, if you can remember that kind of core concept then that goes a long way to carrying these through being able to *reason through a solution*, where if I just know, I just have some equation memorized...that's gonna fall apart, you know, when you get to a situation where that equation doesn't exactly apply. [italic added for emphasis]

This emphasis on reasoning or sense-making from foundational concepts is consistent with the engineering instructor's choice to devote one class period a week to activity around ARS questions.

The interpretation of the different use of the ARS questions by the two instructors is consistent with the analysis of intended thinking processes from the think-aloud interview (Table 2) and the data of percent of correct responses from students (Fig. 5). The biology instructor used the ARS as a periodic check-in with students whereas the engineering instructor used ARS questions more extensively as an opportunity for students to develop their understanding and "create that knowledge structure" they needed for adaptive transfer and problem solving.

**GIW questions** As we found with the ARS questions, instructors also utilized guided inquiry worksheets differently. During the interview, we asked the biology instructor why she recommend the specific guided inquiry worksheet shown in Fig. 2 for the think-aloud interview. She explained that she thought it was the epitome of a typical GIW for the course.

> Biology Instructor (post think-aloud GIW): And so what I really like about POGIL that … this worksheet adheres to is you can get everything you need from this, you know, strictly from the models, and your brain, and *thinking about things*. And maybe if you don't know what these vessels are, yeah, you could look them up, but you probably do, you know, based on where my students are at. And so, like that's what I like. This is very much a standalone.

Here, she expresses how the inquiry-based worksheets in the biology course are designed to be self-contained; there is less emphasis to connecting to the information presented in previous lectures or other places and more emphasis to sense-making or as she says, "thinking about things."

In the engineering course, the GIWs were used in studio sections where the larger class was broken into smaller sections of around 30 students to work on the worksheets. During the interview, the engineering instructor described the relationship between the GIWs and other aspects of the course and especially how they are tied closely to information introduced in lecture.

> Engineering Instructor (post think-aloud GIW): I view studio as a really scaffolded and supported place for students to have their first experience applying the principles from lecture. So you kind of get all this information and not a lot of chances to engage it in lecture and before you get a blank problem statement from the homework assignment and are left with a blank page [you get a chance] *to walk through the steps or the concepts that are gonna have to be applied as we move forward to homework and exams*. Having it be a place where they've got classmates they can bounce ideas [off of]…

Here, he clarifies that he views the guided inquiry worksheets as a useful step for students in between being shown ways to solve problems in lecture and applying these problem solving methods on homework assignments and exams. The engineering instructor further elaborated how he envisions the GIW tool sitting within the instructional processes in the year 3 reflective interview:

> Engineering instructor (g-post): In these studios where students basically come in and they're working on a worksheet on a problem that's related to things that we've covered in class, it's pretty scaffolded, but there's some open ended components, but they're kind of working together in groups of three kind of independently with support from a T.A. during that time.

As this excerpt indicates, when considering active learning tools, it is useful to consider other important aspects of the instructional system, as we do next.

### Answer to RQ2: In what ways do the intended sense-making processes from the ARS and GIW tools align with the instructors' broader perspectives and beliefs about the instructional system for their courses?

Our answer to Research Question 2 is based on analysis of the year 1 general interview (g-pre) and year 3 general interview (g-post).

### Beliefs about ARS and GIW as active learning tools in instructional systems

In this section, we explore more broadly what the instructors conceive as elements of the instructional system and how the ARS and GIW active learning tools fit within those broader elements. Here, the conceptions of the two instructors generally align.

Table 4 shows category codes for elements of the instructional system and examples of the corresponding instructor beliefs that emerged from analyzing the two reflective interviews with each instructor. The table also provides exemplar excerpts from each instructor. Both instructors expressed the tools provided *Instructional Scaffolding* that helped guide students' learning. The excerpt from the biology instructor indicates how she sees scaffolding from the GIW tool as necessary to provide students "a structure to follow," while the engineering instructor describes the role of each tool to progressively provide students "a learning unit" that created a "cohesive, weekly routine."

Both instructors used language that was consistent with a constructivist perspective of learning; the biology instructor often referred to students "constructing their own knowledge", and several times the engineering instructor indicated that he aimed to help students "develop knowledge structures." In addition, both instructors valued the role of *Social Interactions* in constructing understanding, expecting students "to interact, not just with the content but with each other to make meaning of the content" (biology instructor) and "talking with their group and grappling with the material" (engineering instructor). Both instructors allude to how instructional tools can provide the impetus for students to interact with one another in sense-making processes.

Table 4 Instructional system code categories and examples of corresponding beliefs from biology and engineering instructors

| Code | Biology instructor | Engineering instructor |
| --- | --- | --- |
| Instructional scaffolding | "…with my students is that I have to give them structure [with GIW activities]. If I don't give them structure to follow, they don't know what to do." (g-pre) | "… trying to have a cohesive, kind of weekly routine of content delivery: reinforced conceptual understanding in recitation [with ARS questions], scaffolded application in studio [with GIW questions]. So kind of the idea those two lectures, the recitation, and the studio as being like a learning unit…and then the homework follows that…" (g-post) |
| Social interactions | "I expect that they talk to one another and I expect that they synthesize information from whatever we've talked about earlier that morning with what we've done before." (g-pre)<br>"So every day they're expected to interact, not just with the content but with each other and to make meaning of the content." (g-post) | "I'll just have [GIW] worksheets where it's just things like sketch what you think this, you know, the relationship between these two variables is, or…you know, just doing stuff where they're talking with their group and grappling with the material as opposed to me."(g-post) |
| Formative assessment | "But I'm curious what they do know, then based on that data I will choose to, when we start the next day, amend the plan. If it means that we have to have a clicker question the next time to probe this more thoroughly or if maybe I just got to throw it out there and see what they're thinking and where the misconception might be or why they're answering it the way they're answering it." (g-post) | "You get real feedback [from the ARS tool], so do they understand it? 70% of them do or got the right answer, and 30% don't, and so you know you have your finger on the pulse of the class. You know, you're assessing them closer to when you've cover the material, and you're giving them an opportunity to assess their own learning, and so that, and you're giving them an opportunity to communicate what they've learned to their peers." (g-pre). |
| Summative assessment | "When I entered into graduate school, I began teaching anatomy and physiology, which I think traditionally can be looked at as a very memorization-based discipline for anatomy, but for physiology it's process. And then I started crafting exams and assessments that were, you know, more about how could students predict, could students look at a set of data and then make inferences from it, and I came to realize that they couldn't really do that." (g-pre) | "I think it was an exam question, and you know, some students complaining about it being unfair, you know, we haven't covered this in class or whatever, and then just going through what my rationale was. Like if you understood this, the concept from this application then…you know, I was looking to see if you could transfer it and use it over here." (g-pre) |
| Sense-making processes | "We need to not just be looking at content when we do that. We need to be thinking about what does it mean to think like a biologist? You know, what pieces are being gathered or created here, and how are we gonna further them with this course. Content builds, for sure. But what about the thinking like a biologist?" (g-pre)<br>"When I throw my clicker questions out there I say to them, okay, I'm going to need you to defend your choices after these questions come up, I want to hear from you, so I'm prompting them to be reflective about their learning," (g-post) | "and [answering ARS questions] they develop, I think, confidence and a sense responsibility that, you know, I'm not just going to be told the answer here; I have to I have to figure out what the answer is and I think by instilling that in them through this class and the classes that follow they develop skills that they wouldn't develop if you were in a straight lecture classroom." (g-pre)<br>"And it's kind of engineering problem solving is also, a big part of [CBEE] 211 is just being able to take a lot of information, break it up into the parts and map it to, again, those concepts that are kind of fundamental, and then use that information to come to a solution. I think those are the big things I hope they take away from it." (g-post) |

Both instructors also suggested that the data from ARS questions was useful for *Formative Assessment* to "see what they're [i.e., the students are] thinking and where the misconception might be" (biology instructor). The ARS tool allows the instructors to have her or his "finger on the pulse of the class" (engineering instructor) and give students the "opportunity to assess their own learning" (engineering instructor). The engineering instructor also tied this aspect of the instructional system to *Social Interactions* stating ARS questions give "them an opportunity to communicate what they've learned to their peers."

Both instructors recognized the need to develop disciplinary sense-making aligned with their expressed experiences with *Summative Assessments*. As the biology instructor states, "I started crafting exams and assessments that were, you know, more about how could students predict, could students look at a set of data and then make inferences from it, and I came to realize that they couldn't really do that." This realization motivated her to implement POGIL with GIW in her course to help students develop these skills. Similarly, the engineering instructor recalled a time when he received pushback from students for an exam that was perceived as "unfair." He explained his "rationale" to them as follows: "if you understood this, the concept from this application then…you know, I was looking to see if you could transfer it and use it over here." Importantly, both instructors are holding students accountable for higher-level disciplinary thinking processes when they test students, thus aligning the sense-making processes they seek to develop with the active learning tools to the questions on the exams.

In summary, both instructors value and seek to cultivate *sense-making processes*. The biology instructor describes these processes as "thinking like a biologist" which includes defending answer choices and prompting students to be reflective. The engineering instructor expects students to "figure out what the answer is" by "being able to take a lot of information, break it up into the parts and map it to, again, those concepts that are kind of fundamental, and then use that information to come to a [numerical] solution."

## Discussion

In this study, we investigated how two instructors used ARS and GIW tools to identify and compare the ways that they intended for students to "get to the answer." The data show that while the same active learning tools were used in both courses, the way in which students were being asked to engage in problem solving and sense-making varied. In the biology course, ARS questions were used primarily to "check in" with students to see if they were correctly interpreting the worksheet content (e.g., graphs and models) or to ask students to recall the material recently introduced. In the engineering course, ARS questions asked students to apply the concepts covered in lecture to new scenarios towards improving students' conceptual understanding. These uses reflect the activity structure in each course. The biology course centered on briefly using clickers in almost every class to support instruction (lecture or POGIL). In the engineering course, 1 day and 25% of instruction time was devoted to ARS questions, and the instructor asked students to engage in deeper ways by providing written justification and confidence.

In the biology course, the GIWs were primarily used in stand-alone activities, and most of the information necessary for students to answer the questions was contained within the worksheet. Typically, the information was presented in a context that aligned with a disciplinary model. In the engineering course, the instructor intended for students to reference their lecture notes and rely on their conceptual knowledge of fundamental principles from the previous ARS class session in order to successfully answer the GIW questions. The biology instructor used the worksheets as an opportunity for integrated development of their conceptual reasoning, quantitative reasoning, and metacognitive thinking. On the other hand, the engineering instructor focused primarily on cultivating aspects of quantitative reasoning for problem solving.

In our analysis, we position ARS and GIW as tools that are utilized within instructional systems to produce learning. We have shown that the specific intent of the biology instructor when she uses these tools is very different than the engineering instructor. However, common threads emerged that can be used as ways to consider instruction with active learning tools. Both instructors use these tools to build towards the same basic disciplinary thinking and sense-making processes of conceptual reasoning, quantitative reasoning, and metacognitive thinking. Conceptual reasoning processes that were identified in the think-aloud interviews included intending students to use graphical information to qualitatively explain a situation, relate information to a physical representation, and identify relationships between variables. Quantitative reasoning processes included developing equations to describe phenomena and manipulating equations to reveal the relationship between variables. Metacognitive thinking included considering alternative possible solution strategies and reflecting on the reasonableness of an answer value in relation to a physical system.

Both instructors also clearly intended students to interweave these thinking and sense-making processes. The engineering course design was more sequential where students engaged in conceptual reasoning processes during ARS sessions and then were expected to recall those foundational concepts as they were elicited to quantitatively reason with the GIW activity in studio the following day. The biology course design used "POGIL Fridays" to provide a more integrated active learning experience where conceptual reasoning, quantitative reasoning, and metacognitive thinking were more interlocked.

Both instructors clearly alluded to the value of disciplinary thinking processes in each of their general reflective interviews. However, they did not explicitly identify conceptual reasoning, quantitative reasoning, or metacognitive thinking nor did they appear to make these connections in the post think-aloud interviews when they were more specifically asked about the intent of ARS and GIW tools. Thus, the incorporation of conceptual reasoning, quantitative reasoning, and metacognitive thinking appears to be tacit, even for these experienced and highly regarded instructors. We suggest more direct and explicit emphasis on the ways active learning tools elicit these types of thinking would be beneficial as instructors design activities and integrate them into courses.

### Causes of difference in tool use

Hypothetically, we might ask, "If we put one of these instructors in the other's classroom, how similar would their use of the ARS and GIW tools appear in that different context?" There are several legitimate avenues of inquiry that could be pursued to answer this question. We draw from the extant literature to identify these avenues and assert that considering this complex question from several perspectives is productive.

First, we might consider the instructors' beliefs and knowledge. As the set of responses in Table 4 indicate, both instructors demonstrated learner-centered beliefs oriented towards learning facilitation as opposed to teacher-centered beliefs oriented towards knowledge transmission (Prosser and Trigwell 1993). While they shared common orientations, there could be more subtle differences in their beliefs. Speer (2008) suggests a more fine-grained characterization of an instructor's "collection of beliefs" is needed to connect beliefs to specific instructional design choices. Such characterization could provide information about why differences between these instructors' use of the tools emerged. Alternatively, the instructors' designs may be influenced by their knowledge about an educational innovation. Rogers (2003) identifies three types of knowledge needed to implement an innovative tool: awareness knowledge (that the tool exists), how-to knowledge (how to use the tool), and principles knowledge (what purpose the tool serves). In their interviews, each instructor clearly demonstrated awareness and principles knowledge, but differences in how-to knowledge may have led to different enactment strategies. How-to knowledge can be tied to normative use in the department and in the discipline (Norton et al. 2005). For example, there may be more (or different) access to POGIL workshops in biology than in engineering. Further investigation of the degree that detailed instructor beliefs and how-to knowledge influence the choice and use of active learning tools is warranted.

Second, we might consider the different disciplinary contexts of the courses, i.e., biology vs. engineering. The National Research Council (2012) reports that while there are many common pedagogical approaches across science and engineering, there are also "important differences that reflect differences in their parent disciplines and their histories of development" (p. 3). Schwab (1964) argues that each discipline has a unique "structure" leading to particular ways of thinking. Specifically, he distinguishes between thinking associated with "disciplined knowledge (in biology) over the know-how in the solving of practical problems" (p. 267) in engineering. Ford and Forman (2006) extend this framing to disciplinary practices. Each discipline has a unique set of fundamental and central practices that need to be articulated and incorporated into a classroom activity. These sociocultural practices provide access to disciplinary specific ways of thinking, knowing, and justifying. They state that a central goal of education is that students develop "a grasp of practice" which includes both disciplined knowledge and "know-how" (p. 27). This line of inquiry suggests that investigations are needed to elucidate the productive ways active learning tools can support disciplinary practices and the way those uses can differ amongst STEM disciplines or among courses within a discipline.

Third, we might consider how the active learning tools were situated within each course's schedule and institutional resources. The biology class met only in single large-class sections and used undergraduate learning assistants to support POGIL Fridays. The engineering course had dedicated smaller studio sections which were supported by graduate teaching assistants. These different contexts are largely determined by how each department organized classes and support for teaching and likely take sustained effort for an individual instructor to change. Since each course relied upon pedagogically trained student instructors to engage student groups during the use of GIW tools, one of the instructor's roles was to orchestrate and manage an instructional team. In large courses, productive ways to engage the instructional team can become an integral part of incorporating active learning tools (Seymour 2005). In addition, each of the student instructors brings their own knowledge and beliefs about learning to this work (Gardner and Jones 2011). Coordinated activity within the department, college, or university, such as programmatic professional development of student instructors, can become a valuable resource. Research is needed to better understand the ways these greater organizational structures enable or constrain the use of active learning tools.

### Limitations

This study only examined the practices of two instructors within the same institution. It would be useful to verify the findings with a larger sample of instructors and courses that fit within the criteria of the study. This study focused on the intent of the instructors through think-aloud and reflective interviews triangulated with other data sources. In both courses, students were regularly doing work where they were interacting in small groups. It would be useful to see to what degree students were taking up the thinking and sense-making processes of conceptual reasoning, quantitative reasoning, and metacognitive thinking. This take-up clearly depends on the social aspects of learning, involving interactions between the students themselves and the instructor. It would be useful to examine what types of moves by students promote or short-circuit these sense-making processes amongst the group as well as identifying productive ways for an instructor to intervene to facilitate thinking. Finally, while the same three general intended sense-making processes were identified in both the biology and engineering courses, their manifestation undoubtedly depends on the nature of the specific practices of each discipline. Articulation of the specific ways that practicing biologists and engineers engage in disciplinary sense-making could inform more productive uses of these active learning tools.

## Recommendations

This study has led to the following recommendations for post-secondary instructors seeking to integrate active learning tools into STEM courses:

> Recommendation 1: When transitioning to active learning, it is common to think about instructional choices in terms of "pedagogies" like POGIL or Peer Instruction or active learning "technologies" like clickers. We encourage instructors to think about these choices in terms of pedagogically and technology-based active learning "tools." A tool should serve definite educational purposes that are defined prior to use. As with any type of tool, procedural competence is necessary. However, as illustrated in this study, these tools can be used in several ways and their use can become more sophisticated with time.
>
> Recommendation 2: A tool-based orientation should go beyond procedures and prescriptions for delivery. Active learning tools can cultivate disciplinary thinking and sense-making processes that include conceptual reasoning, quantitative reasoning, and metacognitive thinking. Importantly, these processes can bootstrap one another towards deeper understanding (Veenman 2012; Zimmerman 2000). Thus, in designing activity for students, instructors should consider how to progressively integrate the different types of sense-making processes to support one another towards doing disciplinary work and building disciplinary understanding. Integration can be achieved either through a sequence of activities as the engineering instructor did (i.e., conceptual reasoning with ARS followed by quantitative reasoning with GIW) or within a single activity as the biology instructor did (i.e., conceptual reasoning, quantitative reasoning, and metacognitive thinking with POGIL).
>
> Recommendation 3: Active learning tool use needs to account for course structure and context where deliberate choices support learning goals. The biology instructor enacted POGIL Fridays within a standard MWF lecture schedule. The engineering instructor had a split class on Wednesdays to support use of the ARS tool for conceptual understanding and smaller studio sessions on Thursdays for guided inquiry. Instructors should think about their course structures and, if possible, work with administrators to adapt them for better alignment to the tools that support instructional goals.
>
> Recommendation 4: In using active learning tools to promote disciplinary sense-making, instructors of all levels of experience should take a reflective and iterative view of their *instructional practice*. For example, both instructors studied here were acknowledged by students and their peers as excellent—a characterization that was supported by the interview data. But, even so, they could reflect on ways to possibly shift their activity with active learning tools to better align with learning goals. The biology instructor might push students towards conceptual reasoning with delivery of ARS questions, and the engineering instructor might modify his GIW with more emphasis on conceptual reasoning and metacognitive thinking. Rather than viewing such changes in instruction inherently as a criticism of teaching prowess, instructors should view ongoing adjustments as a characteristic of masterful practice.

#### Acknowledgements
The authors are grateful to RMC Research who conducted, audio-recorded, and transcribed the year 3 general reflective interviews, to Jana Bouwma-Gearhart who provided comments on an early version of the manuscript, and to the two instructors who kindly agreed to allow us insight into their teaching practice.

#### Funding
This work was conducted with support from the National Science Foundation under grant DUE 1347817. Any opinions, findings, and conclusions or recommendations expressed in this material are those of the author and do not necessarily reflect the views of the National Science Foundation.

#### Availability of data and materials
The datasets generated and/or analyzed during the current study are not publicly available because this is still an active project and a public release would violate the terms of our IRB approval. Some parts of the data set are available from the corresponding author on reasonable request.

#### Authors' contributions
All authors made substantial contributions to the article and participated in the drafting of the article. All living authors read and approved the final manuscript.

#### Ethics approval and consent to participate
This study has been approved by the Institutional Review Board (IRB) at the authors' institute (study # 6158).

#### Competing interests
The authors declare that they have no competing interests.

#### Publisher's Note
Springer Nature remains neutral with regard to jurisdictional claims in published maps and institutional affiliations.

#### Author details
[1]School of Chemical, Biological, and Environmental Engineering, Oregon State University, Corvallis, OR 97331, USA. [2]College of Education, Oregon State University, Corvallis, OR 97331, USA.

Koretsky et al. International Journal of STEM Education (2018) 5:18

Page 19 of 20

## References

Abraham, MR, & Renner, JW. (1986). The sequence of learning cycle activities in high school chemistry. *Journal of Research in Science Teaching*, 23(2), 121–143.

Association of American Colleges and Universities. (2005) Liberal Education and America's Promise. Retrieved 19 December 2017, from https://www.aacu.org/leap

Bailey, CP, Minderhout, V, Loertscher, J. (2012). Learning transferable skills in large lecture halls: Implementing a POGIL approach in biochemistry. *Biochemistry and Molecular Biology Education*, 40(1), 1–7.

Bean, J. (2016). Set assignments in an explicit, real-world context. Retrieved 19 December 2017, from https://serc.carleton.edu/sp/library/qr/designing_assignments.html#real

Beatty, ID, Gerace, WJ, Leonard, WJ, Dufresne, RJ. (2006). Designing effective questions for classroom response system teaching. *American Journal of Physics*, 74(1), 31–39.

Blasco-Arcas, L, Buil, I, Hernández-Ortega, B, Sese, FJ. (2013). Using clickers in class. The role of interactivity, active collaborative learning and engagement in learning performance. *Computers & Education*, 62, 102–110.

Bogen, J, & Woodward, J. (1988). Saving the phenomena. *The Philosophical Review*, 97(3), 303–352.

Borrego, M, Froyd, JE, Hall, TS. (2010). Diffusion of engineering education innovations: A survey of awareness and adoption rates in US engineering departments. *Journal of Engineering Education*, 99(3), 185–207.

Boscardin, C, & Penuel, W. (2012). Exploring benefits of audience-response systems on learning: A review of the literature. *Academic Psychiatry*, 36(5), 401–407.

Caldwell, JE. (2007). Clickers in the large classroom: Current research and best-practice tips. *CBE-Life Sciences Education*, 6(1), 9–20.

Campbell, T, Schwarz, C, Windschitl, M. (2016). What we call misconceptions may be necessary stepping-stones toward making sense of the world. *Science and Children*, 53(7), 28.

Castillo-Manzano, JI, Castro-Nuño, M, López-Valpuesta, L, Sanz-Díaz, MT, Yñiguez, R. (2016). Measuring the effect of ARS on academic performance: A global meta-analysis. *Computers & Education*, 96, 109–121.

Chari, DN, Nguyen, HD, Zollman, DA, & Sayre, EC. (2017). Student and instructor framing in upper-division physics. *arXiv preprint arXiv:1704.05103*.

Chi, MT, & Wylie, R. (2014). The ICAP framework: Linking cognitive engagement to active learning outcomes. *Educational Psychologist*, 49(4), 219–243.

Chien, YT, Chang, YH, Chang, CY. (2016). Do we click in the right way? A meta-analytic review of clicker-integrated instruction. *Educational Research Review*, 17, 1–18.

Cobb, P. (1994). Where is the mind? Constructivist and sociocultural perspectives on mathematical development. *Educational Researcher*, 23(7), 13–20.

de Jong, T. (2006). Technological advances in inquiry learning. *Science*, 312, 532–533.

de Jong, T, & Van Joolingen, WR. (1998). Scientific discovery learning with computer simulations of conceptual domains. *Review of Educational Research*, 68(2), 179–201.

Dori, YJ, Mevarech, ZR, Baker, DR (Eds.) (2017). *Cognition, metacognition, and culture in STEM education: Learning, teaching and assessment*, (vol. 24). Chas, Switzerland: Springer International Publishing AG.

Douglas, EP, & Chiu, CC (2009). Use of guided inquiry as an active learning technique in engineering. In *Proceedings of the 2009 research in engineering education symposium*.

Dreyfus, BW, Elby, A, Gupta, A, Sohr, ER. (2017). Mathematical sense-making in quantum mechanics: An initial peek. *Physical Review Physics Education Research*, 13(2), 020141.

Duncan, D (2005). *Clickers in the classroom: How to enhance science teaching using classroom response systems*. New York: Addison Wesley and Benjamin Cummings.

Eberlein, T, Kampmeier, J, Minderhout, V, Moog, RS, Platt, T, Varma-Nelson, P, White, HB. (2008). Pedagogies of engagement in science. *Biochemistry and Molecular Biology Education*, 36(4), 262–273.

Elicker, JD, & McConnell, NL. (2011). Interactive learning in the classroom: Is student response method related to performance? *Teaching of Psychology*, 38(3), 147–150.

Engelbrecht, J, Bergsten, C, Kågesten, O. (2012). Conceptual and procedural approaches to mathematics in the engineering curriculum: Student conceptions and performance. *Journal of Engineering Education*, 101(1), 138–162.

Ericsson, KA (2006). Protocol analysis and expert thought: Concurrent verbalizations of thinking during experts' performance on representative tasks. In *The Cambridge handbook of expertise and expert performance*, (pp. 223–242).

Farrell, JJ, Moog, RS, Spencer, JN. (1999). A guided-inquiry general chemistry course. *Journal of Chemical Education*, 76(4), 570.

Felder, RM, & Brent, R. (2010). The National Effective Teaching Institute: Assessment of impact and implications for faculty development. *Journal of Engineering Education*, 99(2), 121–134.

Fies, C, & Marshall, J. (2006). Classroom response systems: A review of the literature. *Journal of Science Education and Technology*, 15(1), 101–109.

Ford, CL, & Yore, LD (2012). Toward convergence of critical thinking, metacognition, and reflection: Illustrations from natural and social sciences, teacher education, and classroom practice. In A Zohar, YJ Dori (Eds.), *Metacognition in science education*, (pp. 251–271). Dordrecht: Springer-Verlag.

Ford, MJ, & Forman, EA. (2006). Chapter 1: Redefining disciplinary learning in classroom contexts. *Review of Research in Education*, 30(1), 1–32.

Freeman, S, Eddy, SL, McDonough, M, Smith, MK, Okoroafor, N, Jordt, H, Wenderoth, MP. (2014). Active learning increases student performance in science, engineering, and mathematics. *Proceedings of the National Academy of Sciences*, 111(23), 8410–8415.

Gardner, GE, & Jones, MG. (2011). Pedagogical preparation of the science graduate teaching assistant: Challenges and implications. *Science Educator*, 20(2), 31.

Geertz, C (1994). Thick description: Toward an interpretive theory of culture. In *Readings in the philosophy of social science*, (pp. 213–231).

Glass, GV, McGaw, B, Smith, ML (1981). *Meta-analysis in social research*. Beverly Hills: Sage.

Grawe, N. (2016). Developing quantitative reasoning. Retrieved 19 December 2017, from https://serc.carleton.edu/sp/library/qr/index.html

Hake, RR. (1998). Interactive-engagement versus traditional methods: A six-thousand-student survey of mechanics test data for introductory physics courses. *American Journal of Physics*, 66(1), 64–74.

Hanson, D, & Wolfskill, T. (2000). Process workshops—a new model for instruction. *Journal of Chemical Education*, 77(1), 120.

Hora, MT, Oleson, A, Ferrare, JJ (2013). *Teaching dimensions observation protocol (TDOP) user's manual*. Madison: Wisconsin Center for Education Research.

Hunsu, NJ, Adesope, O, Bayly, DJ. (2016). A meta-analysis of the effects of audience response systems (clicker-based technologies) on cognition and affect. *Computers & Education*, 94, 102–119.

Kay, RH, & LeSage, A. (2009). Examining the benefits and challenges of using audience response systems: A review of the literature. *Computers & Education*, 53(3), 819–827.

Keselman, A. (2003). Supporting inquiry learning by promoting normative understanding of multivariable causality. *Journal of Research in Science Teaching*, 40(9), 898–921.

Koretsky, M, Bouwma-Gearhart, J, Brown, S. A., Dick, T., Brubaker-Cole, S. J., Sitomer, A., Quardokus Fisher, K., Risien, J., Little, D. L., Smith, C., & Ivanovitch, J. D. (2015). *Enhancing STEM Education at Oregon State University*. Paper presented at 2015 ASEE Annual Conference and Exposition, Seattle, Washington. https://doi.org/10.18260/p.24002

Koretsky, MD, Brooks, BJ, Higgins, AZ. (2016). Written justifications to multiple-choice concept questions during active learning in class. *International Journal of Science Education*, 38(11), 1747–1765.

Koretsky, MD, Falconer, JL, Brooks, BJ, Gilbuena, DM, Silverstein, DL, Smith, C, Miletic, M. (2014). The AIChE Concept Warehouse: A web-based tool to promote concept-based instruction. *Advances in Engineering Education*, 4(1), 7:1–27.

Kuo, E, Hull, MM, Gupta, A, Elby, A. (2013). How students blend conceptual and formal mathematical reasoning in solving physics problems. *Science Education*, 97(1), 32–57.

Lantz, ME. (2010). The use of 'clickers' in the classroom: Teaching innovation or merely an amusing novelty? *Computers in Human Behavior*, 26(4), 556–561.

Laws, P., Sokoloff, D., and Thornton, R. "Promoting active learning using the results of physics education research." UniServe Science News 13 (1999)

Lehrer, R. (2009). Designing to develop disciplinary dispositions: Modeling natural systems. *American Psychologist*, 64(8), 759.

Lesh, RA, & Doerr, HM (2003). *Beyond constructivism: Models and modeling perspectives on mathematics problem solving, learning, and teaching*. Mahwah, NJ: Lawrence Erlbaum.

Lewis, SE. (2011). Retention and reform: An evaluation of peer-led team learning. *Journal of Chemical Education*, 88(6), 703–707.